\documentclass{ifacconf}

\usepackage{amsmath}
\usepackage{amsfonts}
\usepackage{amssymb}
\usepackage{bm}
\usepackage{graphicx}
\usepackage{natbib}
\usepackage{epstopdf}
\usepackage{soul,color}
\graphicspath {{fig/}}

\let\ifacconfcaptionwidth\captionwidth
\usepackage[caption=false]{subfig}
\let\captionwidth\ifacconfcaptionwidth

\newcommand*{\bbar}[1]{\bar{\bar{#1}}}

\usepackage{tikz}
\usetikzlibrary{arrows, decorations.markings}
\usepackage{verbatim}
\tikzset{myptr/.style={decoration={markings,mark=at position 1 with %
    {\arrow[scale=2,>=stealth]{>}}},postaction={decorate}}}

\usepackage[symbol*]{footmisc}
\DefineFNsymbolsTM{otherfnsymbols}{%
	\textbullet \circ
	\textsection   \mathsection
	\textdaggerdbl \ddagger
	\textbardbl    \|%
	\textparagraph \mathparagraph
}%
\setfnsymbol{otherfnsymbols}

\usepackage[bordercolor=white,backgroundcolor=gray!30,linecolor=black,colorinlistoftodos]{todonotes}

\begin{document}
	\begin{frontmatter}
		\title{Design of a Networked Controller 
			for a Two-Wheeled Inverted 
			Pendulum Robot\thanksref{footnoteinfo}}
		\thanks[footnoteinfo]{This work was funded by 
			Deutsche Forschungsgemeinschaft (DFG) 
			within their priority programme 
			SPP 1914 "Cyber-Physical 
			Networking" (CPN). Grant numbers RA516/12-1, 
			CA595/71, and KE1863/5-1.}
		
		\author[TUB]{Zenit Music} 	
		\author[TUB]{Fabio Molinari}
		\author[TUMarch]{Sebastian Gallenm\"uller}	
		\author[TUMcomm]{Onur Ayan}		
		\author[TUMcomm]{Samuele Zoppi}	
		\author[TUMcomm]{Wolfgang Kellerer}		
		\author[TUMarch]{Georg Carle}		
		\author[TUB]{Thomas Seel}
		\author[TUB,MPI]{J\"org Raisch}
		\address[TUB]{Control Systems Group - Technische Universit\"at Berlin, Germany.}	
		\address[TUMarch]{Chair of Network Architectures and Services - Technische Universit\"at M\"unchen, Germany.}		
		\address[TUMcomm]{Chair of Communication Networks - Technische Universit\"at M\"unchen, Germany.}
		\address[MPI]{Max-Planck-Institut f\"ur Dynamik komplexer technischer Systeme, Germany.}	
	\begin{abstract}
		The topic of this paper is	
		to use 
		an intuitive model-based approach to
		design a networked controller 
		for a recent benchmark scenario. 		
		The benchmark problem is to remotely control a
		two-wheeled inverted pendulum robot via W-LAN
		communication. 
		The robot has to keep a vertical upright
		position.
		Incorporating wireless communication in
		the control loop introduces multiple uncertainties and
		affects system performance and stability. The proposed
		networked control scheme employs model predictive techniques
		and deliberately extends delays in order to make them
		constant and deterministic. The performance of the resulting
		networked control system is evaluated experimentally
		with a predefined benchmarking experiment and is
		compared to local control involving no delays.

	\end{abstract}

\end{frontmatter}
\DefineFNsymbolsTM{reset}{%
		\textasteriskcentered *
		\textdaggerdbl \ddagger
}%
\setfnsymbol{reset}
%
\section{Introduction}	
Advancements in communication and computation technology have led to
the concept of Networked Control Systems (NCSs), cf. \cite{hespanha2007survey}.  
In an NCS, components
are distributed and interact via a communication network. This allows
a considerable increase in flexibility, but also raises many design
challenges (\cite{bemporad2010networked},\cite{walsh2002stability}). In
the case of wireless communication, non-deterministic delays and
packet losses are characteristic phenomena.  
{Delays are traditionally assumed 
shorter than the sampling time, see \cite{nilsson1998stochastic},
or compensated by adopting a 
Model Predictive Control scheme as in \cite{mori2014compensation}.
On the other hand,
packet loss is commonly compensated by either 
employing H-infinity control design, see \cite{ishii2008h},
or gain-scheduling feedback, see \cite{yu2008stabilizability}.
A variety of further methods and approaches have been proposed to address these challenges (cf. \cite{zhang2016survey})}.  

\begin{figure}[htb]
		\begin{tikzpicture}[font=\small]
		\draw (0, 0) node[inner sep=0] {\includegraphics[width=\columnwidth]{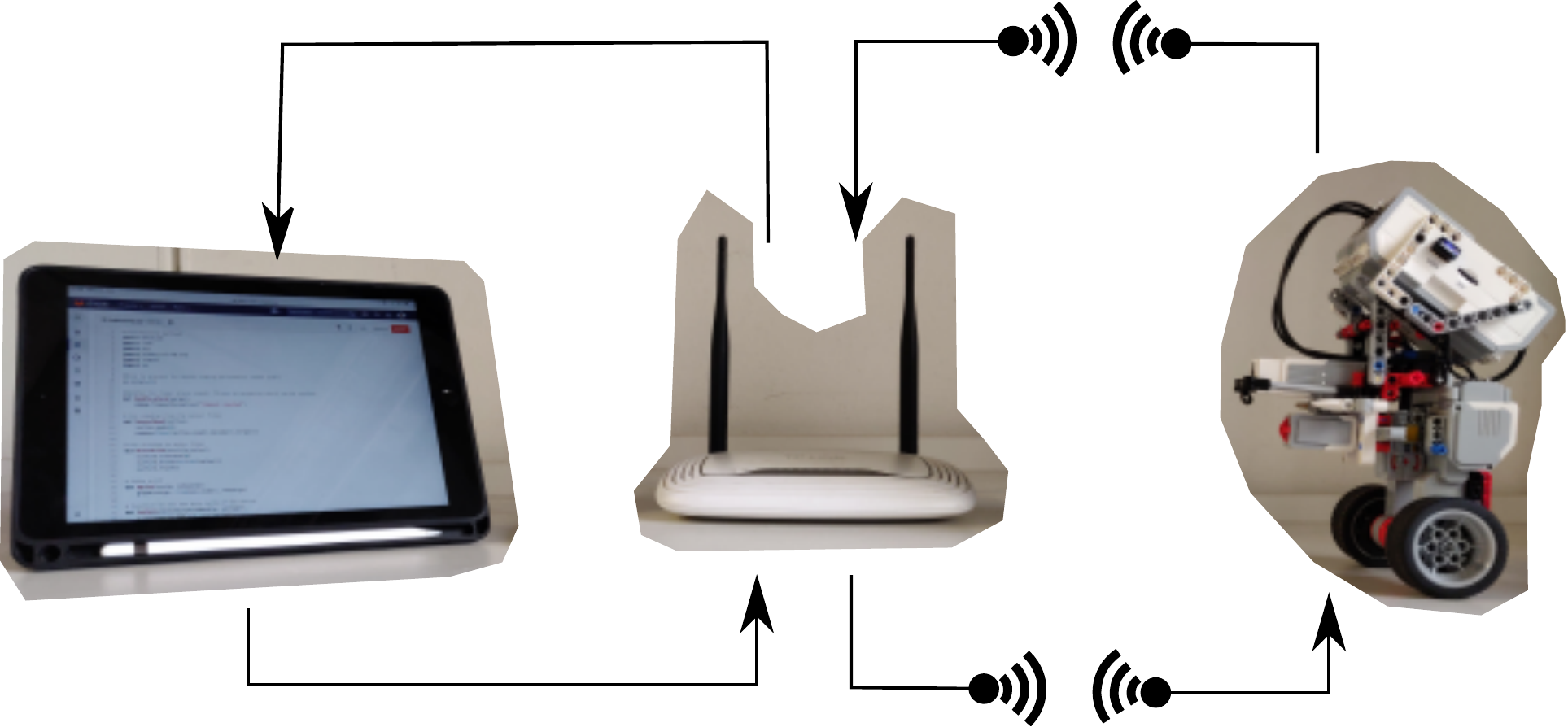}};
		\draw (3.2, -2) node {control signals};
		\draw[densely dotted] (1.5, -1.83)--(1.8, -1.83);
		\draw (1.6, 1.83)--(1.9, 1.83);
		\draw (-1.5, 2) node {measurement signals};
		\end{tikzpicture}
		\caption{Networked Control of a TWIPR.
			In Section~\ref{sec:ncs_arch},
			the asymmetry in wireless data transmission
			(dotted and unbroken lines)
			will be explained.%
                      }
		\label{fig:ncs}
	\end{figure}
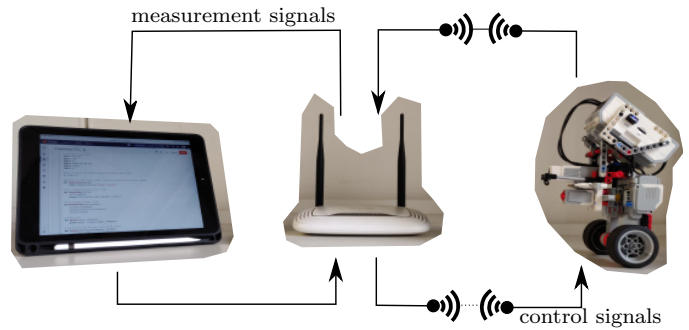

The purpose of this paper is to design a networked controller for a recently proposed
benchmark problem. The benchmark, described in detail in \cite{ipsnpaper} and
\cite{gallenmuller2018benchmarking}, is
to remotely control a two-wheeled inverted pendulum robot (TWIPR)
over
a wireless network, see Fig.~\ref{fig:ncs}. 
To make the benchmark inexpensive
and easily reproducible, the widespread platform LEGO Mindstorm EV3
is used to realize the robot. All details regarding plant and
controller are publicly accessible at {\tt {\small
    https://github.com/tum-lkn/iccps-release}}.

        The body of the robot is mounted on two wheels, on which two DC
        motors are splined.  The control objective is to keep the
        robot in a vertical upright position while undergoing a predefined
        benchmarking experiment.
        To the best of our knowledge,
        the design of a networked controller for a TWIPR has not been
        addressed yet. 
        Among other authors, \cite{pathak2005velocity} stabilized a TWIPR using local,
        i.e., non-networked, control.  
        \cite{ananyevskiy2017control} proposed
        a control strategy over internet for 
        coordinating oscillations of a group of pendula
        and validated it experimentally
        using LEGO Mindstorm NXT.
        Synchronization was not always achieved,
        and the controller did not
        compensate for lost communication packets.
        
        In the current paper we design a networked controller for a TWIPR by using two previously suggested	 strategies 
        (cf. \cite{hespanha2007survey}): 
        (i) we deliberately extend the actuation delays of the 
        NCS in order to make them constant and
        deterministic;
        (ii) a sequence of future control inputs is computed
        from a sequence of model-based state predictions and is sent to the robot
        via the wireless network. If a control packet is lost, the
        robot uses the last received sequence of control inputs
        and applies the input corresponding to the current instant.

          {The remainder of this paper is organized as follows: in
          Section~\ref{sec:prob_desc}, a discrete-time linear
          dynamical plant model is presented and the available sensor
          information is described.  Section~\ref{sec:loc_ctrl}
          introduces  a local controller that is used as a baseline for
          the networked controller developed in Section~\ref{sec:wireless_ctrl}.  The performance of both
          controllers is compared in Section~\ref{sec:performance}.}
	

	In the following, 
	the set of nonnegative (positive) integers is $\mathbb{N}_0$ ($\mathbb{N}$).
	The set of real numbers is $\mathbb{R}$.
	The set of nonnegative (positive) real numbers is denoted 
	$\mathbb{R}_{\geq0}$ ($\mathbb{R}_{>0}$).
	Given a time-dependent real-valued 
	signal  $x:\mathbb{R}_{\geq0}\mapsto\mathbb{R}$,
	its first and second derivative with respect to time are denoted $\dot{x}$ and $\ddot{x}$.	
	Given a vector $\bm{v}\in\mathbb{R}^n$, $n\in\mathbb{N}$,
	its transpose is $\bm{v}'$,
	while the element in position $i\in\{1,\dots,n\}$
	is $\bm{v}_i$. 
	Given a matrix $A\in\mathbb{R}^{n\times m}$,
	its i-th column, with $i\in\{1,\dots,m\}$,
	is $[A]_i$.
	
\begin{figure}
	\subfloat[Side view]{
		\centering
		\includegraphics[width=.4\linewidth]{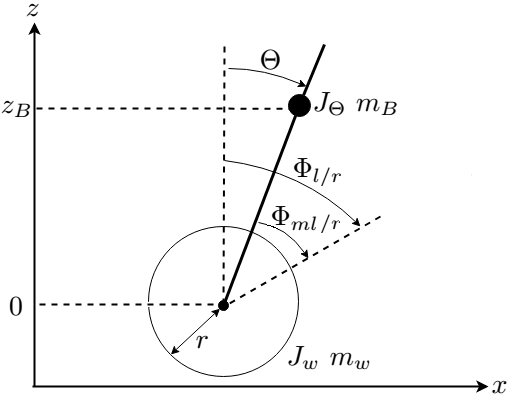}
		\label{fig:roboside}
	}
	\qquad
	\subfloat[Top view]{
		\centering
		\includegraphics[width=.4\linewidth]{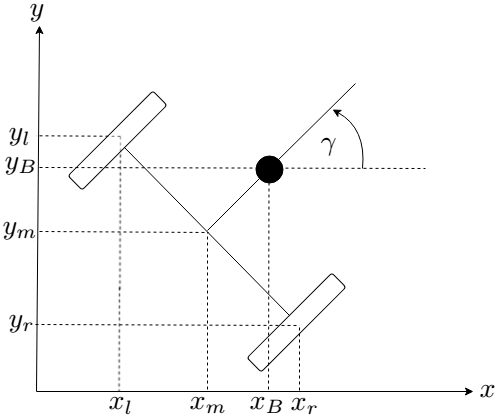}
		\label{fig:robotop}
	}
	\caption{Idealized mechanical model of the TWIPR.}
	\label{fig:robo}
\end{figure}

\section{Problem Description}
	\label{sec:prob_desc}
\begin{table}[t]
	\begin{center}
		\begin{tabular}{l  l p{0.6\linewidth} }			
			Symbol & Unit &  Description \\
			\hline
			$\Theta$& rad & Body pitch angle\\
			$\Phi_{r} (\Phi_{l})$  & rad & Rotation angle of the right (left) wheel\\
			$\Phi_{mr} (\Phi_{ml})$  & rad & Rotation angle of the right (left) motor\\
			$\Phi:=\frac{(\Phi_{r} +\Phi_{l})}{2}$  & rad & Average rotation angle of the wheels\\
			$\gamma$& rad & Yaw angle of the body
		\end{tabular}	
		\vspace{5px}
	\end{center}
	\caption{Kinematic Variables}\label{tab:vars}
\end{table}
	\subsection{Dynamical Model}
		Fig. \ref{fig:robo} sketches
		the robot from the side and the top views
		and illustrates the sign convention;
		the depicted kinematic variables
		are given in Table \ref{tab:vars}.	
		Throughout this section, 
		the focus is not 
		on the derivation of a dynamic model
		for the system, since this has been
		exhaustively addressed elsewhere,
		e.g. the nonlinear dynamic model of a TWIPR
		can be found in \cite[pg. 33]{kim2005dynamic}%
		%
		%
		\makeatletter
		\ifdefined\r@appendA
		\ and in Appendix \ref{appendA}
		of the extended version of this paper%
		\else
		\fi.
		With the model of the TWIPR
		at hand \cite[pg. 33]{kim2005dynamic},
		let
		\begin{align}
			{\bm x}(t) = [ \Phi(t),\Theta(t),\dot{\Phi}(t),\dot{\Theta}(t),\gamma(t),\dot{\gamma}(t)]', 
		\end{align}
		be the state vector and		
			${\bm u}(t)=[u_l(t),u_r(t)]'$
		the input, where 
		$u_{l}(t)$ ($u_r(t)$) is the voltage applied to
		the left (right) DC-motor at time $t$.
		The nonlinear continuous-time model is
		\begin{equation}
			\label{eq:nonlinDynFunc_sec}
			\forall t \in \mathbb{R}_{\geq 0}, \text{ } \dot{{\bm x}}(t) = f({\bm x}(t), {\bm u}(t)),
		\end{equation}	
		and the linearized continuous-time
		dynamics (around the origin) is	
		\begin{equation}
			\label{eq:contlinedynamics_sec}
			\dot{\bm x}(t) = A {\bm x}(t) + B{\bm u}(t).
		\end{equation}
		Matrices $A\in\mathbb{R}^{6\times6}$
		and $B\in\mathbb{R}^{6\times2}$
		are given in \cite[pg. 38]{kim2005dynamic}%
		%
		%
		\makeatletter
		\ifdefined\r@appendA
		\ and in Appendix \ref{appendA}
		of the extended version of this paper%
		\else
		\fi.
		Since sensors and controller
		are implemented based on a digital scheme, the system is discretized using the Forward 
		Euler Method with a discretization step 
		$T_s \in \mathbb{R}_{>0}$,
		which yields the discrete-time linear system,
		$
		\forall k\in\mathbb{N}_0$,
		\begin{equation}
			\label{eq:disclinedynamics_aug_sec}
			{\bm x}_d(k+1) = A_d {\bm x}_d(k) + B_d{\bm u}_d(k),
		\end{equation}
			where $A_d \in \mathbb{R}^{6\times6}$, $B_d \in \mathbb{R}^{6\times2}$ and 
		\begin{equation*}
			{\bm x}_d(k) = {\bm x}(kT_s), \quad {\bm u}_d(k) = {\bm u}(kT_s), \quad \forall k \in \mathbb{N}_0.
		\end{equation*}		
	Finally, for the specific hardware under consideration, a backlash occurs between each motor shaft and the respective wheel (see \cite{nordin}), which is not captured by the model. 
	We define a nonlinear function $\bm{f}_{\mathrm{bl}}:\mathbb{R}^2\mapsto\mathbb{R}^2$ 
	to model the impact of the backlash on the input to the motors.
	Then, (\ref{eq:disclinedynamics_aug_sec})
	is rewritten as 
	\begin{equation}
		\label{eq:disclinedynamics_aug_sec_bl}
		{\bm x}_d(k+1) = A_d {\bm x}_d(k) + B_d\bm{f}_{\mathrm{bl}}({\bm u}_d(k)).
	\end{equation}
	Further details are in \cite[pg. 55]{nordin}.

\subsection{Sensors and Measurements}
\label{sec:sensors}
The robot is equipped with a single-axis 
digital gyroscope (mounted on the body)
measuring the 
body pitch rate. Two digital encoders are
splined on the two motor shafts and
measure the left and the right motor angles.
Let, $\forall k\in\mathbb{N}_0,\ \dot{\Theta}^{(m)}(k)$
be the measurement coming from the gyroscope, and let $b(k)$ be the measurement bias of that gyroscope, which is determined from quasi-static measurements using standard bias estimation approaches, see \cite{tin2011review}. Finally, let $k\in\mathbb{N}_0, \Phi_{ml}^{(m)}(k),\Phi_{mr}^{(m)}(k)$
be the left and right encoder measurements, respectively.


The states of system (\ref{eq:disclinedynamics_aug_sec})
can be estimated as follows: 
\begin{align}
\forall k\in\mathbb{N}_0,\quad 
&{\dot{\Theta}}(k) = \dot{\Theta}^{(m)}(k) - b(k),\label{eq:gyroRate}\\
&{\Theta}(k) ={\Theta}(k-1) + T_s{\dot{\Theta}}(k),\label{eq:est_pitch_angle}\\
&{\Phi}(k) = \frac{\Phi_{ml}^{(m)}(k)+\Phi_{mr}^{(m)}(k)}{2}+{\Theta}(k),\\
&{\dot{\Phi}}(k) = \frac{{\Phi}(k) - {\Phi}(k-1)}{T_s},\\
&{\gamma}(k) = \frac{r}{W}(\Phi_{mr}^{(m)}(k)-\Phi_{ml}^{(m)}(k)),\\
&{\dot{\gamma}}(k) =  \frac{{\gamma}(k) - {\gamma}(k-1)}{T_s}\label{eq:yawRate},
\end{align}
where ${\Theta}(0) = \Theta_0 \in \mathbb{R}$, 
${\rho}(0)= 0$, 
${\Phi}(0)= 0$,
and
${\gamma}(0)= 0$. 
In the following, 
in order to avoid confusion between measured
states and variables,
the measured state vector at time $k\in\mathbb{N}_0$
will be denoted by $\bbar{\bm x}(k)$.
\section{Local Controller}
\begin{figure}
	\includegraphics[width = \columnwidth, trim={.5cm 0 .5cm 0}]{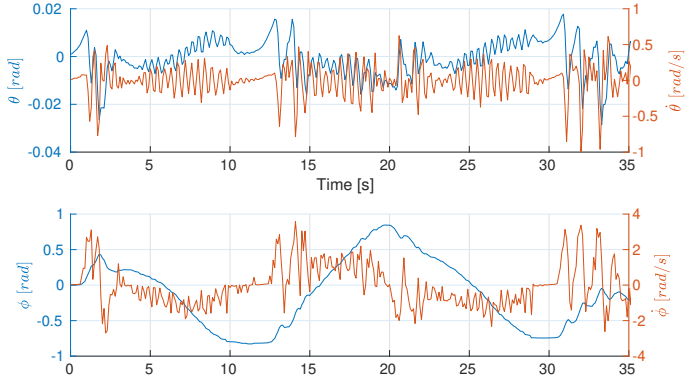}
	\caption{
		Measured states of the
		TWIPR while keeping its upright vertical position
		by employing (\ref{eq:controller}).
	}
	\label{fig:localExp}
\end{figure}
\begin{figure*}[t]
\includegraphics[width = \textwidth, trim={2.2cm 0 2.5cm 0}]{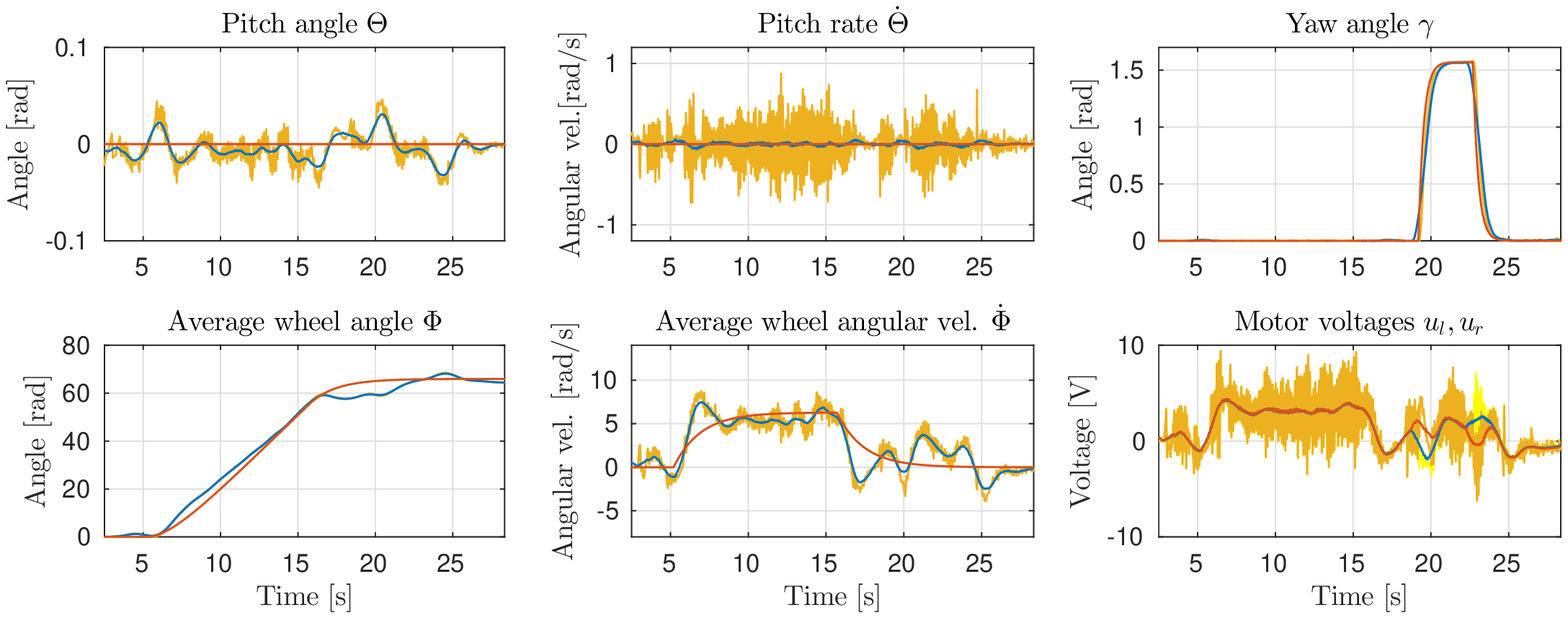}
\caption{Reference tracking experiment using local controller with a sample time of $35$[ms].
	Raw measurements are plotted in yellow, filtered post-processed values in blue, and
	reference in red. 
	The local controller guarantees stability and
	reference tracking, despite the noisy measurements.
}
\label{fig:localRun}
\end{figure*}
	\label{sec:loc_ctrl}
	\subsection{Controller Design}
		\label{seq:controller}
		Due to limited available input voltages,
		we choose a linear quadratic regulator (LQR) design, which allows us to balance 
		between performance and control effort, 
		see \cite{astrom2010feedback}. 
		The feedback control law uses the estimated state from Section~\ref{sec:sensors},
		is based on (\ref{eq:disclinedynamics_aug_sec}), and has the form,
		$
		\forall k\in\mathbb{N}_0$,
		\begin{equation}
			\label{eq:controller}
			{\bm u}_d(k) = -K \bbar{\bm {x}}(k),
		\end{equation}
		where $K\in\mathbb{R}^{2\times6}$ 
		is the control gain matrix
		obtained by minimizing
		the cost function
		\begin{equation}	
			\label{eq:criterion}
			J=\sum_{k=0}^\infty {\bm x}_d'(k)Q{\bm x}_d(k)+ {\bm u}_d'(k)R{\bm u}_d(k),
		\end{equation}
		with $Q\in\mathbb{R}^{6\times6}$ and $R\in\mathbb{R}^{2\times2}$
		given\footnote{$
			R = \mathrm{diag}(10^4, 10^4)
			$, $
			Q = \mathrm{diag}(1, 10^3, 1, 1, 10^6, 1)
			$}.
		The control gain $K$ is obtained 
		by solving the Algebraic Riccati Equation 
		associated to (\ref{eq:criterion}), 
		see  \cite[p. 171]{lalo2014advanced}.
		This controller is responsible for keeping the 
		TWIPR in its upright pose despite
		disturbances. 
		\subsection{Benchmarking Experiment}
		\label{seq:localExperiment}
		In order to facilitate the comparison of 
		different control strategies,
		benchmarking experiments are defined. 		
		The robot body is lifted manually, and 
		measurements (\ref{eq:gyroRate})--(\ref{eq:yawRate}) are started. 
		As soon as the measured pitch angle 
		reaches a neighborhood of $0$,
		say at discrete-time step $\bar{k}\in\mathbb{N}$ 
		corresponding to time $\bar{t}\in\mathbb{R}$,
		the loop is closed and 
		(\ref{eq:controller}) is applied for all $k\geq\bar{k}$. 
		Results from one trial are presented in Figure \ref{fig:localExp}. 
		The local controller assures that all states remain close to zero. 
		
		In order to test a more dynamic and challenging scenario, 
		we also implement the tracking control law
		\begin{equation}			
			\label{eq:controller_wrong}
			{\bm u}(k) = -K( \bbar{\bm{x}}(k)-{\bm x}_\text{ref}(k)),
		\end{equation}
		where ${\bm x}_\text{ref}(k)$ is a given reference state trajectory.
		We use the same gain $K$ as for the stabilitation problem, 
		although we should note that it is no longer an optimal feedback gain 
		for the reference tracking task \cite[pg. 247]{anderson1972linear}.
		In order to obtain smooth and approximately realizable reference trajectories ${\bm x}_\text{ref}(k)$,
		we choose low-pass filtered step changes for $\dot{\Phi}_\text{ref}(k)$ and $\gamma_\text{ref}(k)$, as 
		illustrated in Figure~\ref{fig:localRun}, and determine ${\Phi}_\text{ref}(k)$ and $\dot{\gamma}_\text{ref}(k)$
		by integration and differentiation, respectively.
		Finally, $\theta_\text{ref}(k)$ and $\dot{\theta}_\text{ref}(k)$
		are chosen constantly zero. Note that a non-zero pitch angle trajectory is required
		if the robot accelerates and decelerates. However, 
		for the slowly varying velocity reference 
		signals considered in this contribution, 
		zero is a close approximation of that trajectory.

		Results from an exemplary run of the experiment
		are depicted in Figure \ref{fig:localRun}. 
		
	
		The sampling time of $35[\mathrm{ms}]$ could seem
		large. However, it is required by
		the employed platform,
		which does not
		exhibit state-of-the-art performance.
		Moreover, future work will study
		groups of TWIPRs
		communicating over the channel with orthogonal channel
		access methods, e.g. TDMA (Time Division Multiple Access),
		for which this large sampling time is required.
		
%
%

\section{Wireless Controller}
\label{sec:wireless_ctrl}
\subsection{Networked Control System Architecture}
\label{sec:ncs_arch}
A NCS is mainly composed of three parts,
as illustrated in {Figure \ref{fig:ncs}}:
(i) the plant, in this case the TWIPR, 
equipped with sensors and actuators.
It typically has limited computational power,
which is often not enough for
hosting the controller on-board; 
(ii) the communication network,
in this case a wireless interface
employing the {W-LAN protocol};
(iii) the controller, in this case executed on
a computer with higher computational power.

Data transmission uses
W-LAN according to IEEE\,802.11g (WiFi)
with a transfer rate of 54\,Mbit/s 
in infrastructure mode.
Both robot and controller are directly 
connected to the wireless access point, 
the controller via the Ethernet adapter Intel I219-LM, 
the Robot via the wireless dongle Edimax EW-7811Un.
The WiFi operates indoor in the 2.4\,GHz ISM band, 
which is shared between multiple wireless 
technologies such as other WiFi standards and Bluetooth,
and employs UDP as a transport protocol.
A realistic 
interfered office environment is used, where 
several neighboring WiFi networks are coexisting.
In this environment, 
we observed random packet losses due to "short-time" link failures, 
which could
not be compensated on the communication layer.
\cite{ipsnpaper} contains an exhaustive
statistical analysis of
communication delays for this experimental setup.

At this point, two considerations are derived:
(i) experimentally, 
no packets
sent by the robot
are lost;
this means that the random phenomenon of packet loss
occurs only to packets sent by the controller
(this asymmetry can be seen in Fig. \ref{fig:ncs});
(ii) all transmitted packets
eventually reaching the robot with
delays above a given threshold are
considered as lost.
\begin{figure}[h]
	\resizebox{\columnwidth}{!}{
	\begin{tikzpicture}[font=\huge]
	\node (Robot timeline) at (2,1.5) {TWIPR};
	\node (Host timeline) at (2,-0.9) {Controller};
	\draw [->](2.5,1) to (21,1);
	\draw [->](2.5,-0.4) to (21,-0.4);
	\draw [dashed] (1.5,1) to (2.5,1);
	\draw [dashed] (1.5,-0.4) to (2.5,-0.4);
	\draw (4, 0.8) to (4,1.2);
	\node at (4, 0.3) {$t_{m}^{k}$};
	
	\draw (6, 0.8) to (6,1.2);
	\node[text=gray] at (6, 0.3) {$t_{sr}^{k}$};
	
	\draw (10, -0.2) to (10,-0.6);
	\node at (10, -1.1) {$t_{rh}^{k}$};
	
	\draw (12, -0.2) to (12,-0.6);
	\node[text=gray] at (12, -1.1) {$t_{sh}^{k}$};
	
	\draw (15, 0.8) to (15,1.2);
	\node at (15, 0.3) {$t_{rr}^{k}$};

	\draw (18, 0.8) to (18,1.2);
	\node at (18, 0.3) {$t_{a}^{k}$};
	
	\draw (20, 0.8) to (20,1.2);
	\node at (20, 0.3) {$t_{m}^{k+1}$};

	\draw [dotted] (4,1.2) to (4,2.2);
	\draw [dotted] (6,1.2) to (6,2.2);
	\draw [<->, dashed] (4,2.2) to (6,2.2);
	\node[text=gray] at (5,2.7) {$d_{c1}^{k}$};
	\draw [dotted] (10,-0.2) to (10,2.2);
	\draw [<->, dashed] (6,2.2) to (10,2.2);
	\node[text=gray] at (8,2.7) {$d_{rh}^{k}$};
	\draw [dotted] (12,-0.2) to (12,2.2);
	\draw [<->, dashed] (10,2.2) to (12,2.2);
	\node[text=gray] at (11,2.7) {$d_{c2}^{k}$};
	\draw [dotted] (15,1.2) to (15,2.2);
	\draw [<->, dashed] (12,2.2) to (15,2.2);
	\node[text=gray] at (13.5,2.7) {$d_{hr}^{k}$};
	\draw [dotted] (18,1.2) to (18,2.2);
	\draw [<->, dashed] (15,2.2) to (18,2.2);
	\node at (16.5,2.7) {$d_{c3}^{k}$};
	\draw [dotted] (20,1.2) to (20,2.2);
	\draw [<->, dashed] (18,2.2) to (20,2.2);
	\node[text=gray] at (19,2.7) {$t_{idle}^{k}$};
	
	\draw [dotted] (4,2.2) to (4,3.2);
	\draw [dotted] (20,1.2) to (20,3.2);
	\draw [<->, dashed] (4,3.2) to (20,3.2);
	\node at (12,3.8) {$T_{s}$};
	
	\draw [dotted] (4,2.2) to (4,4.5);
	\draw [dotted] (18,1.2) to (18,4.5);
	\draw [<->, dashed] (4,4.5) -- node[above] {Actuation delay} (18,4.5);
	\end{tikzpicture}
}
	\caption{Timeline of one control cycle of the NCS. 
		All involved delays and times 
		are analyzed in 
		\cite{ipsnpaper}.}
	\label{fig:timeline}
\end{figure}
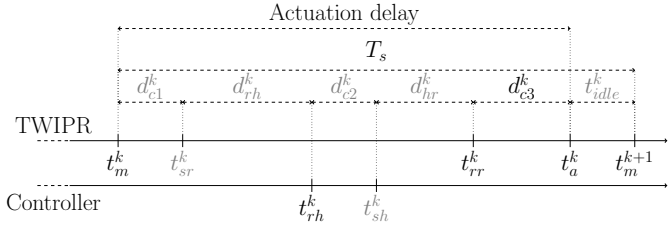

\subsection{Timing Scheme} 
\label{seq:timing}
%
Figure \ref{fig:timeline} shows a timeline 
of the k-th control cycle. 
The following times and
delays between these steps
will be of major relevance in the upcoming sections:
%

\begin{itemize}
	\item $t_m^k$: 
	the robot reads 
	from the sensors. In this moment,
	the control 
	cycle starts;
	
	
	
	
	\item $t_{rh}^k$: 
	the controller receives
	the transmitted packet;
	
	
%
	
	\item$t_{rr}^k$: 
	the robot receives the control input;
	
	\item $d_{c3}^{k}$:
	time 
	between data reception and actuation;
	
	\item $t_{a}^k$: 
	instant when the robot 
	applies the desired input to the motors;
	$t_a^k-t_m^k$ is the \textbf{actuation delay}.
	
\end{itemize}
From a control perspective,
besides dropped packets
from controller to robot,
also \textit{actuation delay}
must be taken into account, 
i.e. the control input
applied at $t_a^k$
is computed based on data measured at $t_m^k$.
This actuation delay is
time-variant, since, in general,
$t_a^{k_1}-t_m^{k_1}\not=t_a^{k_2}-t_m^{k_2}$,
$k_1,k_2\in\mathbb{N}_0$.


\subsection{Wireless Controller Design}
\label{sec:wirConDes}
Control packets reach the robot
before an arbitrary timeout $\tau_o\in\mathbb{R}$ if
%
\begin{equation}
t_{rr}^{k} - t_{m}^{k} < \tau_o < T_s.
\label{eq:delayLeq}
\end{equation}
In the following, 
if (\ref{eq:delayLeq})
does not hold, the packet is acknowledged as lost.
Define a boolean variable that indicates packet loss:
\begin{equation}
\forall k \in \mathbb{N}_0,\ 
\epsilon(k) = 
\begin{cases} 
1 &\text{if }t_{rr}^{k} - t_{m}^{k} \geq \tau_o\\
0 &\text{ otherwise}
\end{cases} .
\end{equation} 
So, packets are lost either 
because never delivered (trivial) or
because they would arrive too late at the receiver.
To avoid non-deterministic uncertainties 
resulting from the time-variance 
of the actuation delay, 
we deliberately dilate $d_{c3}^k$ 
(by putting a waiting time
before the processed signal
is applied
to the motors)
such that 
\begin{equation}
\label{eq:strategyCombatTVD}
\forall k \in \mathbb{N}_0,\ 
d_{c3}^{k} = T_s - (t_{rr}^k - t_{m}^{k}).
\end{equation}
In any $k\in\mathbb{N}_0$
where (\ref{eq:delayLeq}) holds,
by (\ref{eq:strategyCombatTVD}), 
$d_{c3}^{k}>0$.
This strategy leads to a larger but constant actuation delay.

We compensate this delay by model-based prediction. 
Define the following inputs, $\forall k\in\mathbb{N}_0$:
\begin{align*}
&{{\bm u}}_{id}(k) := -K{\bm x}(t_{a}^{k}),\qquad
&\hat{{\bm u}}(k) := -K \hat{{\bm x}}(t_{a}^{k}),
\end{align*}
where 
${{\bm u}}_{id}(k)$ is the ideal control input 
based on states at time of actuation and
$\hat{{\bm u}}(k)$ is a control input, 
computed based on a state $\hat{{\bm x}}(t_{a}^{k})$ that is predicted
from the last available measurement 
at $t_m^k$, 
i.e. $\bbar{\bm{x}}(k)$. 
The state is predicted by integrating 
the nonlinear dynamics (\ref{eq:nonlinDynFunc_sec}):
\begin{equation}
\label{eq:nonlinsim}
\hat{{\bm x}}(t_{a}^{k}) = \bbar{\bm x}({k})+ \int\limits_{t_{m}^{k}}^{t_{m}^{k}+T_s} f({\bm x}(t), {\bm u}(t_{m}^{k})) \mathrm{dt}.
\end{equation}
In case (\ref{eq:delayLeq}) does not hold,
we use model-based prediction to compensate packet loss. 
The controller
does not know \textit{a-priori}
whether a packet will be lost. Therefore, it 
calculates and sends a list of $M+1$ control inputs, 
which are computed based on model-based state predictions of the next $M+1$ steps, where $M$ is chosen larger than the expected maximum number of packet losses. 
Formally, $\forall k \in \mathbb{N}_0$, 
the controller computes and sends the control input matrix
\begin{equation}
\label{eq:controlMat}
\hat{U}(k) := 
\left[ \hat{{\bm u}}(k), {\hat{{\bm u}}}(k+1), \dots, {\hat{{\bm u}}}(k+M)\right] 
\in \mathbb{R}^{2\times M+1},
\end{equation}
where, $\forall k \in \mathbb{N}_0,\
\forall i \in \{1,\dots,M\}$,
\begin{equation*}
	{\hat{{\bm u}}}(k+i) = -K{\hat{{\bm x}}}(t_a^{k+i}),
\end{equation*}
with, $\forall k \in \mathbb{N}_0,\
\forall i \in \{0,\dots,M-1\}$,
\begin{equation}
	\label{eq:predictionMat}
	{\hat{{\bm x}}}(t_a^{k+i+1}) =
	{\hat{{\bm x}}}(t_a^{k+i}) + 
	\int\limits_{t_{m}^{k+i}}^{t_{m}^{k+i}+T_s} f({\bm x}(t), \hat{\bm u}({k+i})) \mathrm{dt}.
\end{equation}
If the computation time at the remote
controller is such that (\ref{eq:delayLeq}) is in general violated,
predictions (\ref{eq:nonlinsim}) and (\ref{eq:predictionMat})
can be computed using the
linearized discrete-time system (\ref{eq:disclinedynamics_aug_sec_bl}),
which is computationally faster but less accurate. The control matrix sent to the robot would, then, be
\begin{equation*}
	{\hat{U}}_l(k) = \left[ {\hat{{\bm u}}}_l(k), {\hat{{\bm u}}}_l(k+1), \dots, {\hat{{\bm u}}}_l(k+M)\right] \in \mathbb{R}^{2\times M+1},
\end{equation*}
where, $\forall k \in \mathbb{N}_0,\ \forall i \in \{0,\dots,M\},$
\begin{equation}
	\hat{{{\bm u}}}_l(k+i) = -K{\hat{{\bm x}}}_l(k+i),
\end{equation}
with
\begin{multline}
	{\hat{{\bm x}}}_l(k+i) = A_d{\hat{{\bm x}}}_l(k+i-1)+ B_d\bm{f}_{\mathrm{bl}}
	\left({\hat{{\bm u}}}_l(k+i-1)\right)
\end{multline}
and
${\hat{{\bm x}}}_l(k-1) = \bbar{\bm x}({k})$, 
${\hat{{\bm u}}}_l(k-1) = {\bm u}_d({k})$.
%
Then, the most recent 
control matrix is
\begin{equation}
U^*(k) := 
\begin{cases} 
{\hat{U}}_l(k) & \text{ if } \epsilon(k) = 0\\
U^*(k-1) & \text{ otherwise} 
\end{cases},
\end{equation}
and the TWIPR applies the control input
\begin{equation}
\forall k \in \mathbb{N}_0,\;{\bm u}(k) = \left[U^*(k)\right]_{\omega(k)},
\end{equation} 
with 
\begin{equation}
\omega(k) := 1 + \min_{\substack{l  \in \left\{0,\dots,M\right\}:\\
		\epsilon(k-l) = 0}} l
\end{equation}
being the number of sampling periods that have passed since the last control matrix was received.

%
The described Networked Control System is implemented with predictions based on the linearized model (\ref{eq:disclinedynamics_aug_sec_bl}). 

\subsection{Experimental Evaluation of the Networked Controller}
\begin{figure}
	\includegraphics[width = \columnwidth,trim={1.5cm 0 .7cm 0}]{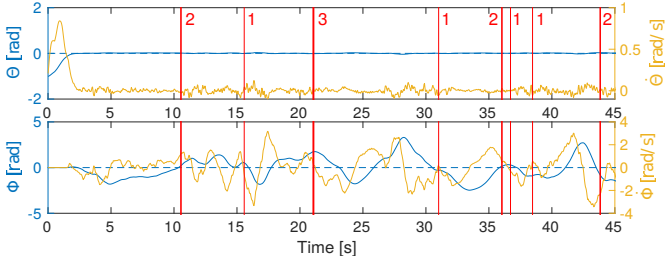}
	\caption{
		Each red vertical line represents an occurrence of packet loss
		(the number indicates the amount of sequential packet loss).
		Although at $t=21[s]$
		three control signal packets are lost,
		the robot does not lose stability.
	}
	\label{fig:packet_loss}
\end{figure}
First, we
analyze stability
of the proposed NCS and robustness
with respect to
packet losses. Results are given in Figure \ref{fig:packet_loss}.
Just as the local controller, the networked controller assures that the states remain stable but with clearly larger deviations from zero, at least for the wheel angle. 

The proposed networked control strategy
is found to exhibit
robustness against up to 3 sequential packet losses,
i.e.\ the controller can maintain stability for at least $140[\mathrm{ms}]$ without measurement updates.

\begin{figure*}
	\includegraphics[width = \textwidth, trim={2.2cm 0 2.5cm 0}]{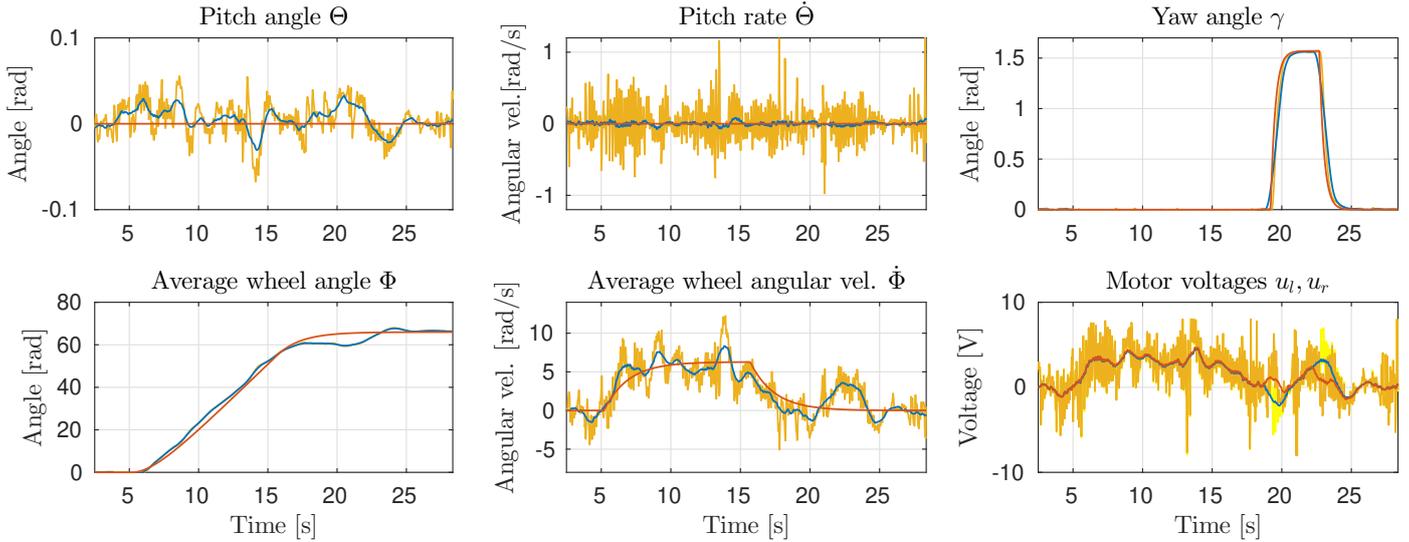}
	\caption{Reference tracking experiment using the NCS with a sampling time of $35$[ms]. 
		Raw measurements are plotted in yellow, filtered post-processed values in blue, and
		references in red.
		The networked control strategy
		exhibits tracks the given references while guaranteeing stability,
		despite the presence of network delays,
		packet loss, and measurement noise.
	}
	\label{fig:wifiRun}
\end{figure*}

\begin{figure}
	\includegraphics[width = \columnwidth, trim={0.2cm 0 .5cm 0}]{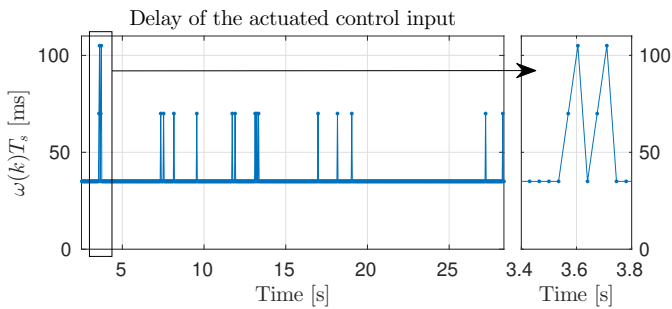}
	\caption{
		Actuation delay, i.e. $\omega(k)T_s$, of every control input.
	}
	\label{fig:input_delay}
\end{figure}

The reference tracking experiment described in Section~\ref{seq:localExperiment} is repeated with the proposed networked controller.
Result of one trial are given in Figures~\ref{fig:wifiRun} and~\ref{fig:input_delay}. 
Despite two consecutive double packet loss
at $t\approx 3.5[\mathrm{s}]$,
measurement noise,
and actuation delay of $35[\mathrm{ms}]$,
the robot does not lose stability.

\section{Performance Comparison}
\label{sec:performance}
We compare the performance of the local and the networked controller for the reference tracking experiment. The local controller has a negligible delay, whereas the networked controller has an actuation delay of $35[\mathrm{ms}]$ (up to $105[\mathrm{ms}]$ in case of packet loss). Performance
degradation due to these delays is evaluated by three \textit{root mean squared error} (RMSE) indices defined by
{
	\begin{align*}
	&\mathrm{RMSE}_\Phi = \sqrt{\frac{1}{k_{\mathrm{end}}-k_0}\sum_{k=k_0}^{k_{{\mathrm{end}}}} (\bbar{\Phi}(k) - \Phi_{ref}(k) )^2},\\
	&\mathrm{RMSE}_\Theta =\sqrt{ \frac{1}{k_{{\mathrm{end}}}-k_0}\sum_{k=k_0}^{k_{{\mathrm{end}}}} (\bbar{\Theta}(k))^2},\\
	&\mathrm{RMSE}_\gamma = \sqrt{\frac{1}{k_{{\mathrm{end}}}-k_0}\sum_{k=k_0}^{k_{{\mathrm{end}}}} (\bbar{\gamma}(k) - \gamma_{ref}(k) )^2},
	\end{align*}
}%
where $k_0,k_{end} \in \mathbb{N}$. 
These RMSE indices
are averaged over 
{a set of ten} 
trials. The results are given in Table \ref{tab:rmse_values}.   
While differences in $\Phi$- and $\gamma$-movement 
errors are marginal, RMSE$_\Theta$ is almost twice as large for the networked controller than for the local controller, although still small in magnitude.

\begin{table}[h]
	\centering
	\begin{tabular}{l | c | c }			
		& Local  &  NCS\\
		\hline
		$RMSE_\Phi$& 2.4141 [rad] & 2.4512 [rad] \\
		$RMSE_\Theta$  & 0.0116 [rad] & 0.0192 [rad]\\
		$RMSE_\gamma$  &  0.0849 [rad] & 0.0888 [rad]\\
	\end{tabular}

	\caption{{RMSE} for the states $\Phi$, $\Theta$ and $\gamma$; values for both the local controller and the NCS. }
	\label{tab:rmse_values}
\end{table}
\section{Conclusion}
\label{sec:conclusion}
In this work, common
networked control strategies have been
implemented
for stabilizing a TWIPR, built with an inexpensive and
highly reproducible platform (LEGO Mindstorm EV3).
The networked controller is able to stabilize the
TWIPR over a wireless channel despite 
time-varying delays and packet loss.
Benchmark experiments have been defined, 
so that local and remote controllers
could be compared.

Future work will consider both
practical and theoretical advancements:
on one hand, custom hardware is being developed.
On the other hand, 
aiming at improving performance shown in Section \ref{sec:performance},
both a state observer and 
an optimal control strategy for reference tracking
(computationally more expensive than
the solution here adopted)
will be employed. 

%

 	
\bibliography{bibliography}

\end{document}